\documentclass[a4paper,fleqn,usenatbib]{mnras}
\usepackage[T1]{fontenc}
\usepackage{ae,aecompl}
\usepackage{graphicx}
\usepackage{amsmath}
\usepackage{amssymb}
\usepackage{natbib}
\usepackage{enumitem}
\usepackage[utf8]{inputenc}
\usepackage{epstopdf}

\title[Kinematically hot young, nearby stars]{A kinematically hot population of young stars in the solar neighbourhood}

\author[A. S. Binks et al.]{
A. S. Binks,$^{1}$\thanks{E-mail: a.s.binks1@keele.ac.uk}
R. D. Jeffries$^{1}$
and N. J. Wright$^{1}$
\\
$^{1}$Astrophysics Group, School of Chemistry and Physics, Keele University, UK}
\date{Accepted XXX. Received YYY; in original form ZZZ}
\pubyear{2020}

\begin{document}
\label{firstpage}
\pagerange{\pageref{firstpage}--\pageref{lastpage}}
\maketitle

\begin{abstract}
In the last three decades several hundred nearby members of young stellar moving groups (MGs) have been identified, but there has been less systematic effort to quantify or characterise young stars that do not belong to previously identified MGs. Using a kinematically unbiased sample of 225 lithium-rich stars within 100 pc, we find that only $50 \pm 10$ per cent of young ($\lesssim 125$ Myr), low-mass ($0.5<M/M_{\odot}<1.0$) stars, are kinematically associated with known MGs. Whilst we find some evidence that six of the non-MG stars may be connected with the Lower Centaurus-Crux association, the rest form a kinematically ``hotter'' population, much more broadly dispersed in velocity, and with no obvious concentrations in space. The mass distributions of the MG members and non-MG stars is similar, but the non-MG stars may be older on average. We briefly discuss several explanations for the origin of the non-MG population.
\end{abstract}


\section{Introduction}\label{sec:intro}

It has been known for some decades that many young ($<125$ Myr) stars exist in the solar neighbourhood, that are not part of obvious star clusters or associations. These can be identified simply from their early spectral-types for high-mass stars \citep[e.g.][]{1983a_Eggen} or from various youth indicators like enhanced magnetic activity, rapid rotation or lack of lithium depletion in the case of low-mass stars \citep[e.g.][]{1990a_Soderblom,1995a_Favata,1995a_Jeffries}. It has become recognised that many of these young stars are members of several unbound, but kinematically coherent ``moving groups'' (MGs) of $10-10^2$ members, with spatial extents of a few tens of pc and velocity dispersions less than a few ${\rm km\,s}^{-1}$ \citep[e.g.][]{2001a_Montes,2001a_Zuckerman,2004a_Zuckerman}. The origins of these MGs are still uncertain, but their youth and proximity means their members have become key targets for investigating exoplanets, discs and sub-stellar objects \citep[e.g.][]{2004a_Kalas,2009a_Lagrange,2015a_Bowler,2015a_MacGregor,2015a_Chauvin} and their study should lead to insights into the mechanisms by which stars leave their (presumably) clustered birth environments and disperse into the field \citep[e.g.][]{2018a_Wright}.

Young, nearby stars can be found using two broad methodologies. The first isolates stars with astrometry matching those of known MGs. Spectroscopy is subsequently required to confirm youth and determine if the age and 3D kinematics are consistent with the MG \citep[e.g.][]{2012a_Schlieder,2013a_Malo,2014a_Gagne,2017a_Riedel,2018a_Gagne}. Such methods successfully find new members of existing MGs but preclude identification of young stars that are not members of known MGs.

The second approach uses large, kinematically unbiased, catalogues of stars with some indication of stellar youth (e.g. X-ray activity). Follow-up spectroscopy is used to measure line-of-sight kinematics and confirm youth. Whilst this method is often less efficient (e.g. old, tidally locked binaries can masquerade as fast rotating, active young stars), it can successfully identify young stars irrespective of their kinematics \citep[e.g,][]{2006a_Torres,2009a_Guillout,2018a_Frasca}.

Recent years have seen several large-scale spectroscopic surveys undertaken to look for nearby, young, low-mass ($<1 M_{\odot}$) stars, free from kinematic bias in their selection process. In these surveys the most useful spectroscopic indicator of youth is the strength of the Li~{\sc i} feature at 6708\AA, since lithium is partly or fully depleted during the pre main sequence phase in stars below a solar mass \citep{2014a_Jeffries,2016a_Lyubimkov}. By comparing the equivalent width (EW(Li)) of the line with that seen in stars of similar spectral-type in clusters of known age, substantial age discrimination can be achieved. Several such surveys have recently found that many young stars cannot be associated with known MGs \citep[e.g.][]{2015a_Binks,2017a_Kastner,2020a_Binks}.

In this work we compile positions, proper motions and parallaxes (and often radial velocities) from the second {\it Gaia} data release \citep[][herein GDR2]{2018a_Gaia_Collaboration} for Li-rich stars identified in previous kinematically unbiased searches for young stars. This sample is used to investigate the kinematics of young stars within 100 pc of the Sun, with the aim of quantifying what fraction of these stars belong to known MGs, searching for any evidence of new MGs and shedding light on the processes by which young stars disperse into the field from their birth environments.

\section{Input Catalogue}\label{sec:catalogue}

Our sample is built chiefly from kinematically unbiased searches for young stars among X-ray selected samples that have EW(Li) measurements (not upper limits). We include 1470 such stars from the ``Search for Associations Containing Young Stars'' (SACY; Torres et al. 2006) and 612 from the RasTyc catalogue \citep{2009a_Guillout, 2018a_Frasca}. The SACY and RasTyc samples were derived from ROSAT X-ray source catalogs matched with the Tycho-2 catalogue \citep{2000a_Hog} and then observed spectroscopically. SACY contains young stellar candidates from a full range of RA, mostly in the southern hemisphere, whilst RasTyc is focused on the northern hemisphere between 8h $<$ RA $<$ 15h. The northern hemisphere sample is supplemented by four smaller, kinematically unbiased, samples described in \cite{2015a_Binks, 2018a_Binks} (216 stars with EW(Li) that were initially selected on the basis of short rotation periods), \cite{2019a_Schneider} and \cite{2019a_Bowler} (28 and 30 stars with EW(Li) that were selected on the basis of UV or X-ray activity alone, respectively). 

This initial selection of 2349 stars was cross-matched with GDR2 and we filtered out any stars that did not have a 5-parameter astrometric solution or had parallax uncertainties $>20$ per cent, or a parallax $>10$ mas (i.e. beyond 100 pc). We have confirmed that the inclusion or not of objects that would fail the GDR2 astrometric quality checks discussed by \cite{2018a_Lindegren} does not change the overall results and conclusions of this paper. The GDR2 astrometric quality checks are that an object must have (i) a renormalised unit weighted error \citep[RUWE, see][]{2018b_Lindegren}~is $<1.4$ and (ii) the number of visibility periods used in the GDR2 astrometric solution is $\geq 8$.

The limiting magnitude of the ROSAT/Tycho-2 samples is imposed by sensitivity limits from the Tycho data ($V \sim 12$). At a distance of 100 pc, the latest spectral-type observable is approximately K5; the sampled volume is therefore smaller for young stellar candidates that are cooler than this. The supplementary sample from \cite{2015a_Binks,2018a_Binks} are limited by the sensitivity limits of the SuperWASP All-Sky Survey and ROSAT ($V \sim 13$), whereas the samples from \cite{2019a_Bowler} and \cite{2019a_Schneider} are sensitive to fainter stars, but are aimed at identifying a volume-limited sample of young M dwarfs. For inclusion in our sample both a GDR2 $G$ magnitude, 2MASS \citep{2003a_Cutri} $K_{\rm s}$ magnitude and at least one RV measurement is required. We obtained supplementary RV measurements by searching the VizieR database\footnote{\url{http://vizier.u-strasbg.fr}} and employed the method described in \cite{2020a_Binks} to calculate an average RV and remove any stars with $> 1$ measurement that have significantly varying RVs that indicate binarity.

\begin{figure*}
\centering
\includegraphics[width=0.90\textwidth,angle=0]{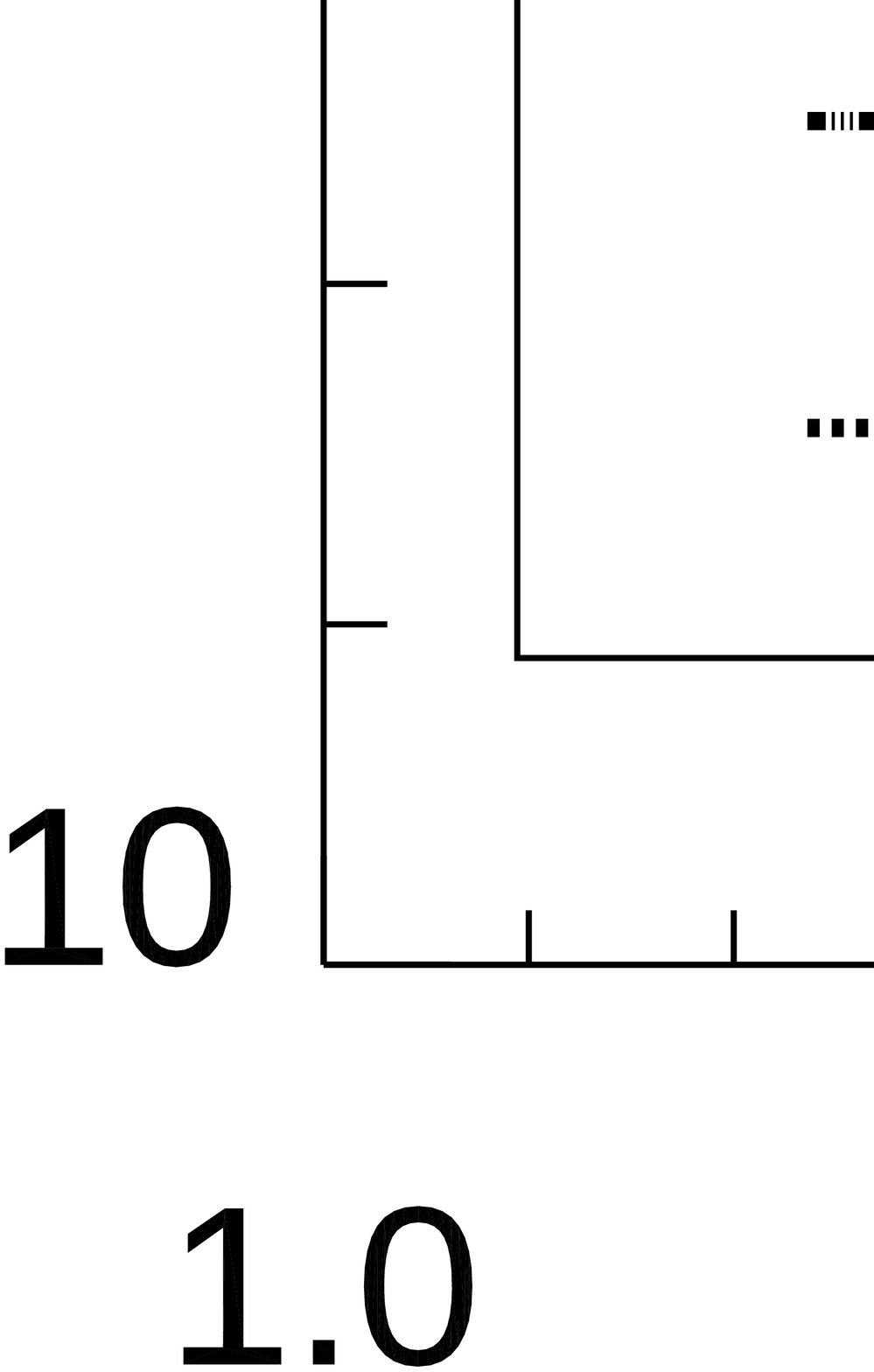}
\caption{Top-left (1a): EW(Li) versus $G-K_{\rm s}$ distribution for our NLYS. Objects in red represent stars that have no kinematic match with any MG and blue objects are stars that have at least one kinematic match (see $\S$\ref{sec:kinematics}). Filled red squares represent 5 stars that are concentrated in velocity space, but not associated with any known MGs within 100 pc (see $\S$\ref{sec:new_groups}). Filled circles represent NLYS that have a renormalised unit weight error (RUWE, see technical note: GAIA-C3-TN-LU-LL-124-01) $< 1.4$ and more than 8 visibility periods in GDR2; and open circles fail at least one of these criteria. Squares and triangles represent members of NGC 2264 (5 Myr) and the Pleiades, respectively. Top-right (1b): The spatial projection of NLYS in celestial coordinates, where the outlined shapes represent the boundaries of the the Scorpius-Centaurus association \citep[defined in][]{1999a_de_Zeeuw}, which comprises of three sub-regions: Upper Scorpius (USCO), Upper Centaurus Lupus (UCL) and the Lower Centaurus Crux (LCC), the Taurus star forming region \citep[defined in][]{2017a_Esplin} and the Pleiades open cluster \citep[whose boundaries are estimated from the middle-left panel of figure 1 in][]{2019a_Lodieu}. Bottom-left (1c): Absolute $G$ versus $G-K_{\rm s}$ CMD of the NLYS with PARSEC theoretical isochrones at 10 Myr, 80 Myr and 2 Gyr \citep{2017a_Marigo}. Bottom-right (1d): Age and mass distributions calculated using SED fitting, where the lines delimit regions discussed in $\S$\ref{sec:kinematics}, and triangles denote lower age limits as described in $\S$\ref{sec:catalogue}.}
\label{figure:plot}
\end{figure*}

To identify genuinely young stars, Figure~\ref{figure:plot}a compares the EW(Li) and $G-K_{\rm s}$ of stars in our sample (we assumed zero reddening for these nearby stars) with the distribution observed in the Pleiades cluster \citep[$\sim 125$ Myr,][dereddened with $E(G-K_{\rm s})=0.08$]{1998a_Stauffer,2018a_Bouvier}. To be classified as ``young'', a star must have $G-K_{\rm s} \ge 1.4$, corresponding to spectral-types of early-G or later, and an EW(Li) that is at or above the median EW(Li) in the Pleiades at the same intrinsic colour. The latter is approximated by a quartic polynomial fit to a 5-element running median of EW(Li). For cooler stars ($G-K>3.3$), there is no Li detected in the Pleiades but it is present in younger clusters (see below). For these stars we demand that EW(Li)  exceeds three times its measurement uncertainty or 50 m\AA\ if no uncertainties are given. For comparison Figure~\ref{figure:plot}a also shows EW(Li) versus $G-K_{\rm s}$ for a population of very young stars in NGC 2264 \citep[age $\sim 5$ Myr,][]{2016a_Bouvier}.

Applying these criteria results in a catalogue of 225 nearby, likely young stars (NLYS). Details of the sample are listed in Table~\ref{table:input_data} (only available electronically), including the positions, EW(Li), colours and magnitudes, astrometry and RV measurements. The majority of the $\sim 90$ per cent of stars eliminated in our selection process have insufficient EW(Li). The majority (167) of the NLYS are found in the southern hemisphere.

The ages and masses of the NLYS (given in Table~\ref{table:input_data}) were estimated by fitting spectral energy distributions using GDR2, UCAC4 \citep{2012a_Zacharias}, 2MASS photometry and GDR2 parallaxes, using the evolutionary models of \cite{2017a_Marigo} and the method described in \cite{2019a_Wright}. The ages derived from our SED fitting incorporate theoretical isochrones up to ages of 200 Myr, but the SED fitting method becomes insensitive to age when stars approach, or reach, the zero age main sequence (ZAMS). There are 57 NLYS with positions on the CMD close to the main sequence (see Figure~\ref{figure:plot}c), causing their posterior age upper limits to be unconstrained. We used a set of isochrones with $\log{\rm age (yr)} = 7.0~(0.1)~8.0$ and estimated the $G-K_{\rm s}$ colour where each isochrone intersects a $\log {\rm age (yr)} = 9.0$ isochrone.  If the 84$^{\rm th}$ percentile age from the SED fit effectively overlaps with the ZAMS then the age for this star is quoted as a lower limit using the 16$^{\rm th}$ percentile of the posterior age distribution. These limits are denoted with triangles in Figure~\ref{figure:plot}d.

{\centering
\begin{table}
\caption{Input data description for the 225 NLYS. This table is available only in electronic format.}
\begin{center}
\begin{tabular}{lll}
\hline
\hline
Label & Units & Description \\
\hline
2MASSJ                  &                & 2MASS Identifier                         \\
$\pi$                   & mas            & Parallax                                 \\
$\sigma_{\pi}$          & mas            & Uncertainty in $\pi$                     \\
$\mu_{\alpha}$          & mas\,yr$^{-1}$ & Proper-motion in right ascension         \\
$\sigma_{\mu_{\alpha}}$ & mas\,yr$^{-1}$ & Uncertainty in $\mu_{\alpha}$            \\
$\mu_{\delta}$          & mas\,yr$^{-1}$ & Proper-motion in declination             \\
$\sigma_{\mu_{\delta}}$ & mas\,yr$^{-1}$ & Uncertainty in $\mu_{\delta}$            \\
$G$                     & mag            & GDR2 apparent $G$ magnitude              \\
$\sigma_{G}$            & mag            & Uncertainty in $G$                       \\
$K_{\rm s}$             & mag            & 2MASS apparent $K_{\rm s}$ magnitude     \\
$\sigma_{K_{\rm s}}$    & mag            & Uncertainty in $K_{\rm s}$               \\
EW(Li)                  & m\AA           & EW of Li-feature at 6708\AA              \\
$\sigma_{\rm EW(Li)}$   & m\AA           & Uncertainty in EW(Li)                    \\
$M$                     & $M_{\odot}$    & Stellar mass                             \\
$M_{84}$                & $M_{\odot}$    & 84$^{\rm th}$ percentile $M$             \\
$M_{16}$                & $M_{\odot}$    & 16$^{\rm th}$ percentile $M$             \\
Age type                &                & Lower limit ($>$) or measurement ($=$)   \\
Age                     & Myr            & Stellar age                              \\
Age$_{84}$              & Myr            & 84$^{\rm th}$ percentile Age             \\
Age$_{16}$              & Myr            & 16$^{\rm th}$ percentile Age             \\
$U$                     & km\,s$^{-1}$   & Velocity (Galactic centre)               \\
$\sigma_{U}$            & km\,s$^{-1}$   & Uncertainty in $U$                       \\
$V$                     & km\,s$^{-1}$   & Velocity (Galactic rotation)             \\
$\sigma_{V}$            & km\,s$^{-1}$   & Uncertainty in $V$                       \\
$W$                     & km\,s$^{-1}$   & Velocity (Galactic North Pole)           \\
$\sigma_{W}$            & km\,s$^{-1}$   & Uncertainty in $W$                       \\
$RV_{1}$                & km\,s$^{-1}$   & Radial velocity measurement 1            \\
$\sigma_{RV_{1}}$       & km\,s$^{-1}$   & Uncertainty in $RV_{1}$                  \\
$RV_{2}$                & km\,s$^{-1}$   & Radial velocity measurement 2            \\
$\sigma_{RV_{2}}$       & km\,s$^{-1}$   & Uncertainty in $RV_{2}$                  \\
$RV_{3}$                & km\,s$^{-1}$   & Radial velocity measurement 3            \\
$\sigma_{RV_{3}}$       & km\,s$^{-1}$   & Uncertainty in $RV_{3}$                  \\
$RV_{4}$                & km\,s$^{-1}$   & Radial velocity measurement 4            \\
$\sigma_{RV_{4}}$       & km\,s$^{-1}$   & Uncertainty in $RV_{4}$                  \\
$RV_{5}$                & km\,s$^{-1}$   & Radial velocity measurement 5            \\
$\sigma_{RV_{5}}$       & km\,s$^{-1}$   & Uncertainty in $RV_{5}$                  \\
$RV_{6}$                & km\,s$^{-1}$   & Radial velocity measurement 6            \\
$\sigma_{RV_{6}}$       & km\,s$^{-1}$   & Uncertainty in $RV_{6}$                  \\
$RV_{7}$                & km\,s$^{-1}$   & Radial velocity measurement 7            \\
$\sigma_{RV_{7}}$       & km\,s$^{-1}$   & Uncertainty in $RV_{7}$                  \\
$r_{RV}$                &                & References for RV measurement (1)        \\
$RV_{\rm fin}$          & km\,s$^{-1}$   & Final calculated RV                      \\
$\sigma_{RV_{\rm fin}}$ & km\,s$^{-1}$   & Uncertainty in $RV_{\rm fin}$            \\
Bflg                    &                & Flag on binarity status (2)              \\
$\chi^{2}_{\rm MG}$     &                & MGs with $\chi^{2}_{\rm MG} < 2.60$ (3)  \\
$BAN_{MG}$              &                & Best MG match from BANYAN$\Sigma$        \\
$BAN_{P}$               &                & BANYAN$\Sigma$ membership probability    \\
Pref                    &                & Original survey publication (1)          \\
\hline
\end{tabular}
\begin{flushleft}	
Notes: (1) The referenced publications are: a = \cite{2006a_Bobylev}, b = \cite{2006a_Gontcharov}, c = \cite{2006a_Torres}, d = \cite{2014a_Elliott}, e = \cite{2017a_Kunder}, f = \cite{2018a_Gaia_Collaboration}, g = \cite{2018a_Soubiran}, h = \cite{2014a_Malo}, i = \cite{2017a_Shkolnik}, j = \cite{2018a_Binks}, k = \cite{2009a_Guillout}, l = \cite{2018a_Frasca}, m = \cite{2007a_Kharchenko}, n = \cite{2007a_White}, o = \cite{2015a_Binks}, p = \cite{2017a_Kraus}, q = \cite{2000a_Alcala}, r = \cite{2019a_Schneider}, s = \cite{2013a_Lopez_Marti}, t = \cite{1995a_Reid},  u = \cite{2013a_Kordopatis}, v = \cite{2015a_Luo}, w = \cite{2009a_Lepine}, x = \cite{2019a_Bowler}. (2) Binary flags are described in \cite{2020a_Binks}. (3) Ordered from lowest to highest $\chi^{2}_{\rm MG}$ value.
\end{flushleft}
\label{table:input_data}
\end{center}
\end{table}}

\section{Assigning moving group membership}\label{sec:MG_definition}

The original definition of a ``moving group'' goes back to the works of \citet{1961a_Eggen, 1975a_Eggen, 1983a_Eggen}, who identified a group of young early-type stars, supergiants and young clusters in the solar vicinity, with common space motions and ages of $20-150$ Myr, that was dubbed the ``Local Association''. \citet{1986a_Innis}, \citet{1991a_Anders}, \citet{1994a_Jeffries}, \citet{1995a_Jeffries} and others, found many nearby examples of young solar-type stars that, on the basis of their similar space motions, appeared to be the low-mass counterpart of the Local Association. More recently, the Local Association has been fragmented, both kinematically and spatially into a dozen or more named MGs, and several new MGs have been found with kinematics distinct from the Local Association \citep[e.g][and references therein]{2001a_Zuckerman, 2008a_Torres}.

Assigning membership to a particular MG or any MG is not a trivial task. Some authors have maintained the philosophy of earlier works and simply looked for consistency of space motions between candidate members and the kinematics of objects that are used to define the MGs \citep{2012a_Shkolnik, 2014a_Riedel, 2015a_Binks}. A possible issue here is that if MG members are coeval, are born in a relatively compact configuration, and then drift to their current locations, then not using positional information or taking account of how far a star could have moved during its lifetime could lead to erroneous MG assignments (both false positives and false negatives). 

A second general approach is to use both kinematic and spatial information. The velocities and positions of stars can be compared with ellipsoids that define the MGs in both spatial and velocity coordinates. Examples of this technique include the BANYAN code \citep{2013a_Malo} and its subsequent improvements, culminating in BANYAN$\Sigma$ \citep{2014a_Gagne, 2018a_Gagne}, and the LACEwING code \citep{2017a_Riedel}, which also uses a Galactic potential to trace back the positions of stars using their observed velocities. This approach assumes that members of a MG began their lives together in close spatial proximity (or at least much closer together than they are now) and that the ``bona fide'' members used to define the MG ellipsoids are representative of the whole MG population. Whilst there has been some success in tracing back the members of MGs to much smaller volumes in the past \citep{2019a_Crundall}, the former assumption may well be challenged by spectroscopic observations that indicate some degree of chemical inhomogeneity among ``bona fide'' members of the AB Doradus MG \citep{2013a_Barenfeld} and also by recent work identifying nearby ($<500$ pc), kinematically coherent filamentary streams of young stars, that may be several hundreds of pc in length \citep{2019a_Meingast,2019a_Curtis,2020a_Roser,2020a_Beccari}. The latter assumption may also be problematic given that the MG velocity dispersions used in LACEwING and BANYAN$\Sigma$, multiplied by the corresponding MG ages, are usually many times larger than the spatial extents of the defining members. It is therefore probable that some MG members have moved considerably beyond the domain spanned by the defining members, even if they were born in a similar location; or they may be located at considerable distance from the defining members because they originated in a different place.

In this work the main analysis is limited to candidate members of unbound MGs within 100 pc of the Sun. Comparison with known MGs is based initially solely on kinematic criteria (see $\S$\ref{sec:chisq}), using the velocity centroids and dispersions reported in table~7 of \citet{2018a_Gagne}. This analysis is then contrasted with results obtained using the more restrictive kinematic {\it and} spatial constraints adopted by the BANYAN$\Sigma$ code (see $\S$\ref{sec:BANYAN}).

\section{Kinematics of young, nearby stars}\label{sec:kinematics}

Heliocentric Galactic space velocities in the $UVW$ system and their uncertainties are calculated following \cite{1987a_Johnson}. The NLYS are tested for membership of 12 groups with age $<150$ Myr and whose centroids are located within 100 pc, defined in \cite{2018a_Gagne}\footnote{These are all classed as MGs, except the Volans-Carina Association \citep{2018b_Gagne}, which may be more spatially compact.}.

\subsection{Kinematic membership analysis}\label{sec:chisq}

\begin{figure*}
\centering
\includegraphics[width=0.90\textwidth,angle=0]{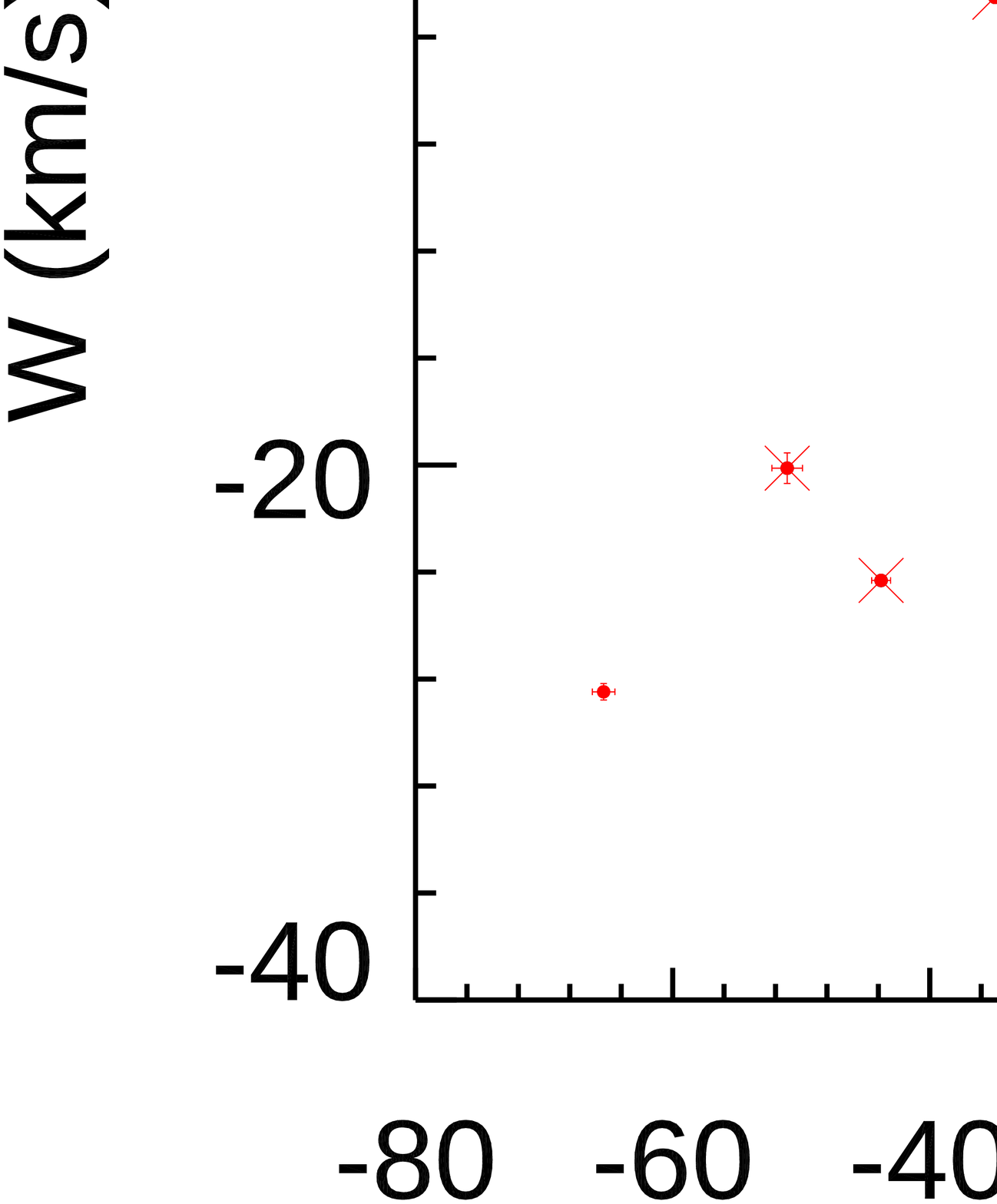}
\caption{$UVW$ velocity distributions for our NLYS, with boxes indicating the $1\sigma$ velocity dispersions for the 12 nearby, young ($<150$ Myr) MGs and associations included in the BANYAN$\Sigma$ code \citep{2018a_Gagne}. Stars denoted with cross-signs have $\geq 2$ consistent RV measurements and are likely single.}
\label{figure:UVW}
\end{figure*}

Velocity uncertainties are combined with the 1$\sigma$ MG velocity dispersions and a chi-squared test of membership is done, using only the velocity information, with a threshold that would reject membership with 95 per cent confidence \citep[e.g, see][]{2012a_Shkolnik, 2015a_Binks, 2018a_Binks}. The result is that 89/225 of NLYS cannot be assigned to any of the MGs. Input data, $UVW$ velocities and $XYZ$ positions, individual RV measurements, and the details of kinematic matches to MGs (if any) are listed in Table~\ref{table:input_data}. The distribution of the sample in $UVW$, compared with the MG velocities is shown in Figure~\ref{figure:UVW}.

Figure~\ref{figure:plot} shows distributions for NLYS that have kinematics consistent with MG membership, versus those that do not, in: the EW(Li)/$G-K_{\rm s}$ plane (the diagram from which the whole sample was selected); sky position (in celestial coordinates); the absolute colour-magnitude diagram; and the age/mass plane. 

Figure~\ref{figure:plot}b shows no obvious concentrations of the non-MG stars; they are distributed across the sky, with some bias towards the southern hemisphere, in a similar way to the likely MG members. There are no significant over-densities towards the nearest star forming regions (SFRs) and young clusters that lie just beyond the horizon of the sample (e.g. Taurus, Pleiades, Sco-Cen). Figure~\ref{figure:plot}c shows the NLYS are found close to, or above the ZAMS, consistent with their Li-rich status. 

Figure~\ref{figure:plot}d shows masses and ages determined from the SED modelling. The overall morphology of this diagram is likely set by the physics of Li depletion and observational selection.  Fully convective low-mass stars deplete their Li quickly, so we do not expect to see Li-rich stars older than $\sim 30$ Myr below $0.5 M_{\odot}$. The $G-K_{\rm s} > 1.4$ colour selection ensures that we also should not include stars $>1 M_{\odot}$ unless they are very young, have low $T_{\rm eff}$ and are still contracting towards the ZAMS. For these reasons, the upper age limit of our Li-selected sample is mass-dependent. Despite these restrictions it appears probable that the proportion of non-MG NLYS increases with age. For stars with $0.5 < M/M_{\odot} < 1.0$ the fraction increases from 20/65 (31 per cent) for ages $<20$ Myr to 49/107 (46 per cent) for older stars, which is inconsistent with the hypothesis of a uniform fraction at a marginal 95 per cent confidence. At lower masses this conclusion also has some independent support from the distribution of EW(Li). Figure~\ref{figure:plot}a shows that at $G-K_{\rm s} > 2.5$, where EW(Li) is strongly age dependent, only 5/39 of the NLYS with EW(Li)$>350$ m\AA, which are likely very young, are not assigned to a known MG. However, we cannot be sure that this fraction increases at older ages, because the sample will be incomplete for low-Li stars. These unassigned Li-rich objects are discussed in more detail in $\S$\ref{sec:Li-rich}.

\begin{figure}
\centering
\includegraphics[width=0.45\textwidth,angle=0]{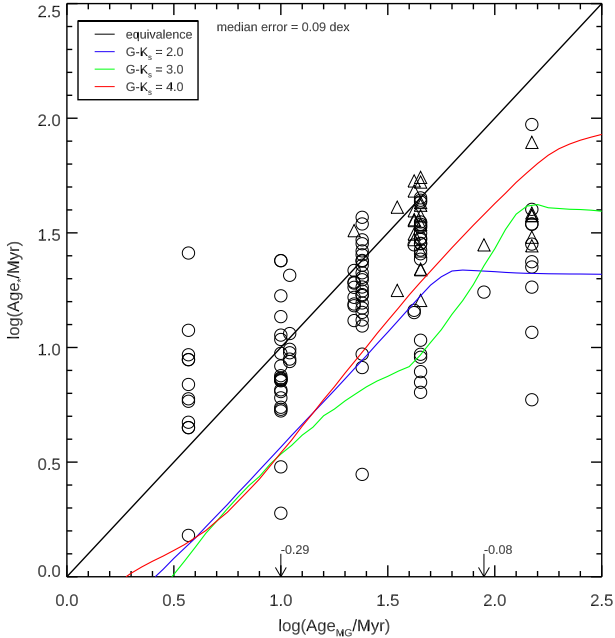}
\caption{Difference between the estimated age from SED fitting and the MG age for 136 NLYS that satisfy kinematic criteria for membership in $\S$\ref{sec:kinematics}. Circles denote NLYS with age measurements from SED fits and triangles are lower age limits (see $\S$\ref{sec:catalogue}). The black line denotes equivalence, while the blue, green and red lines represent (approximately) the younger isochronal ages that would be inferred from SED fitting to equal mass binaries at $G-K_{\rm s} = 2.0, 3.0$ and $4.0$, respectively.}
\label{figure:age_comparisons}
\end{figure}

The 136 NLYS with kinematic matches to at least one MG are compared with the age of the MG in Figure~\ref{figure:age_comparisons}. For NLYS with two or more matches, we choose the MG whose age most closely matches the SED age. The MG comparison ages are from table~1 in \cite{2018a_Gagne}, with the exception of Argus \citep[$45 \pm 5$\,Myr,][]{2019a_Zuckerman}, the Volans-Carina Association \citep[$89^{+5}_{-7}$\,Myr,][]{2018b_Gagne} and the Carina Association \citep[$24$\,Myr,][]{2019a_Schneider}. There is reasonable agreement between the SED ages and the ages of the MG to which they have been assigned, although the scatter is considerably larger than the age uncertainties (the median age uncertainty is $\pm 0.09$ dex). Some of this scatter may be caused by variability (the source photometry was not co-temporal), but there is also a contribution from unresolved binarity -- binary stars will be more luminous and their inferred ages will be underestimates. The extent of the effects of binarity are illustrated by lines in Figure~\ref{figure:age_comparisons}, which show what the inferred age would be for an equal-mass binary at a range of $G-K_{\rm s}$. Most objects are consistent with, or well above these loci, but there is some evidence that a few objects are much younger and may not belong to their assigned MGs {\it if} coevality is a requirement for membership.

%One major contribution to the observed difference between the SED ages and the MG ages might occur if a significant number of these NLYS are in unresolved binary systems. This increases the observed luminosity, and leads to underestimated ages in the SED fitting. To test the effects of unresolved binarity we used the PARSEC v1.2s models to compare the isochronal ages of equal mass binary systems for $G-K_{\rm s} = 2.0, 3.0$ and $4.0$ with the single sequence, denoted as various lines in Figure~\ref{figure:age_comparisons}. The underestimation of our SED ages is consistent with the notion that these could be unresolved multiple components, with the exception of a handful of examples, which may not be members of their assigned MG. The median formal uncertainty in the SED ages is $0.09$ dex, several times lower than the observed age dispersion in each MG (typically $0.4-0.6$ dex). Whilst unresolved binarity ($\sim 0.3$ dex) and variability ($< 0.1$ dex) are contributing factors in the observed age dispersion, this cannot fully encapture the spread in ages, and may indicate that these ensembles are in fact not entirely coeval.}

It is possible that some genuine MG members may have been scattered into the non-MG sample by their velocity uncertainties. Conversely, some non-MG members will have been mistakenly assigned MG membership. This issue is dealt with by assuming that non-MG stars are approximated by uniform distributions in velocity space over the $30^3$(km/s)$^3$ volume shown in the right-hand panels of Figure~\ref{figure:UVW}. The fraction of observed stars that belong to MGs ($f_{\rm obs}$) is corrected to a true fraction ($f_{\rm true}$) using a Monte Carlo simulation  with $10^{5}$ simulated $UVW$ coordinates. Error bars are randomly assigned from the NLYS sample and the fraction of objects $f_{\rm rand}$, that provide at least one match to the MGs, with rejection confidence $P_{\rm th}<0.95$, is calculated. From there an estimate for $f_{\rm true}$ is given by
\begin{equation}
f_{\rm true} = (f_{\rm obs} - f_{\rm rand})/(P_{\rm th} - f_{\rm rand})\, ;
\end{equation}
and we find $f_{\rm true} =0.61 \pm 0.06$.
 
Additional tests of robustness were: varying $P_{\rm th}$ between 0.90 and 0.99; limiting the comparison to stars within 75 pc or 50 pc and excluding stars that do not pass the GDR2 astrometric quality tests (described in $\S$\ref{sec:catalogue}). The results for each of these experiments are presented in Table~\ref{table:experiments}. All experiments are reasonably consistent with the original $f_{\rm true} = 0.61$ within statistical uncertainties. The inclusion of GDR2 astrometric criteria systematically increases $f_{\rm true}$ by $\sim 0.05$, as does considering samples restricted to closer distances. 

If the MG membership fraction is age dependent, then including those stars with $M>1.0 M_{\odot}$ and $M<0.5 M_{\odot}$, where Li-rich objects will be restricted to $\lesssim 30$ Myr, might bias the result. If we restrict the sample just to the range $0.5<M/M_{\odot}<1.0$, where we expect EW(Li) to be more uniformly sensitive to ages $<100$ Myr (see Figure~\ref{figure:plot}d), the fraction of MG members is 103/172, and after correcting for contamination and velocity uncertainties, is still $0.61 \pm 0.06$.

\begin{figure}
\centering
\includegraphics[width=0.45\textwidth,angle=0]{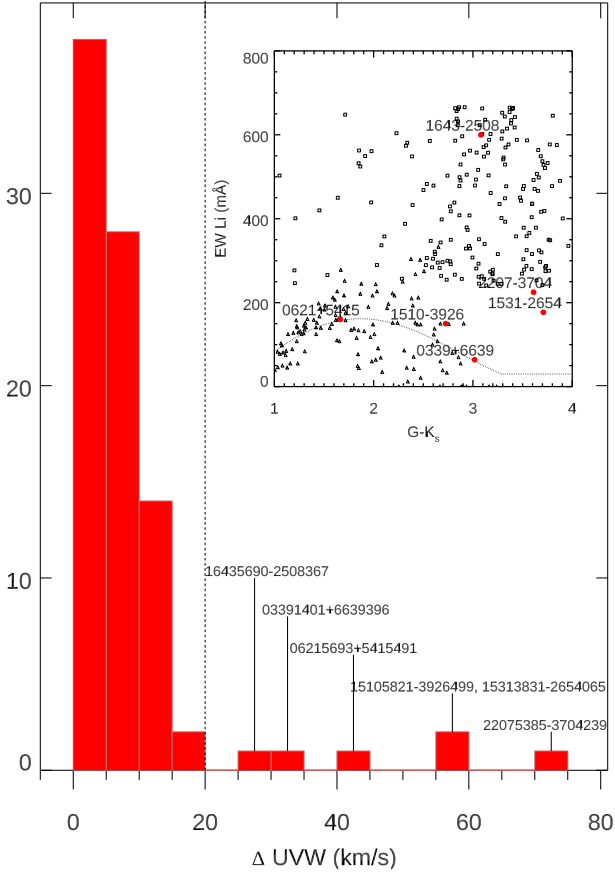}
\caption{Histogram of the 3D velocity separation between objects in the non-MG population to their closest MG in velocity space (within 100 pc). Inset: EW(Li) versus $G-K_{\rm s}$ distribution for 6 stars with $\Delta UVW > 25\,{\rm km\,s}^{-1}$.}
\label{figure:deltaUVW}
\end{figure}

Figure~\ref{figure:UVW} shows that the non-MG NLYS form a kinematically ``hotter'' and more widely dispersed population than those that can be assigned MG membership. There are 22 NLYS whose 3D velocities are $>10\,{\rm km\,s}^{-1}$ from the centroids of the nearest MG in velocity space (i.e. $\Delta UVW =\sqrt{(\Delta U)^2 + (\Delta V)^2 + (\Delta W)^2} >10\,{\rm km\,s}^{-1}$, see Figure~\ref{figure:deltaUVW}). Six of these %(J15105821$-$3926499, J16435690$-$2508367, J03391401+6639396, J06215693+5415491, J15313831$-$2654065 and J22075385$-$3704239) 
have $25 < \Delta UVW < 75\,{\rm km\,s}^{-1}$, of which five are unlikely to have erroneous velocities due to close binarity (see below) since they have at least 2 consistent RV measurements. Three of these have Li EWs compatible with being younger than $\sim 30$ Myr (see Figure~\ref{figure:deltaUVW}, inset). A further object, J15093920$-$1332119, has $\Delta UVW \sim 240\,{\rm km\,s}^{-1}$ but the single RV measurement has very large uncertainties \citep[$-256.16 \pm 42.15\,{\rm km\,s}^{-1}$,][]{2017a_Kunder}~and may be unreliable.

Excluding J15093920$-$1332119, the 3D velocity dispersions of the MG stars and the non-MG NLYS are 7.3 and $19.8\,{\rm km\,s}^{-1}$, respectively. These values are not changed significantly if the samples are restricted to those that pass GDR2 astrometric quality criteria (changing to 6.9 and $21.1\,{\rm km\,s}^{-1}$, respectively). A histogram of $\Delta UVW$ for the non-MG sample is presented in Figure~\ref{figure:deltaUVW}.

{\centering
\begin{table}
\caption{MG member fractions based on our kinematic analysis for different distance limits, membership rejection probability thresholds and GDR2 astrometric quality criteria. $N_{\rm stars}$ gives the number of stars assigned MG membership as a fraction of the total number in the sample. $f_{\rm true}$ is the MG member fraction after correcting for contamination and velocity uncertainties.}
\begin{center}
\begin{tabular}{lrrr}
\hline
\hline
Distance & $P_{\rm th}$ & $N_{\rm stars}$ &  $f_{\rm true}$ \\
limit    (pc) &        &                 &       \\
\hline
\multicolumn{4}{c}{Without GDR2 astrometric criteria} \\
     100 &               0.95 &         136/225 & $0.61 \pm 0.06$ \\
         &               0.99 &         157/225 & $0.68 \pm 0.05$ \\
         &               0.90 &         122/225 & $0.58 \pm 0.04$ \\
      75 &               0.95 &          83/129 & $0.66 \pm 0.06$ \\
         &               0.99 &          93/129 & $0.70 \pm 0.06$ \\
         &               0.90 &          76/129 & $0.64 \pm 0.05$ \\
      50 &               0.95 &           39/59 & $0.68 \pm 0.09$ \\
         &               0.99 &           42/59 & $0.69 \pm 0.09$ \\
         &               0.90 &           36/59 & $0.66 \pm 0.08$ \\
\hline
\multicolumn{4}{c}{With GDR2 astrometric criteria}     \\
     100 &               0.95 &         118/179 & $0.68 \pm 0.05$ \\
         &               0.99 &         134/179 & $0.73 \pm 0.06$ \\
         &               0.90 &         107/179 & $0.65 \pm 0.06$ \\
      75 &               0.95 &          75/107 & $0.72 \pm 0.07$ \\
         &               0.99 &          82/107 & $0.75 \pm 0.07$ \\
         &               0.90 &          70/107 & $0.71 \pm 0.06$ \\
      50 &               0.95 &           35/49 & $0.74 \pm 0.10$ \\
         &               0.99 &           36/49 & $0.72 \pm 0.10$ \\
         &               0.90 &           33/49 & $0.74 \pm 0.10$ \\
\hline
\end{tabular}
%\begin{flushleft}	
%\end{flushleft}
\label{table:experiments}
\end{center}
\end{table}}

The hotter kinematics of the non-MG sample could be caused by unresolved and unrecognised binarity (recall that RV variables have already been excluded). However, there are several arguments that suggest this is unimportant: (i) The fraction of non-MG NLYS with multiple, consistent RV measurements is 77 per cent, suggesting they are not binaries with large velocity amplitudes. (ii) There are 20 non-MG members with just a single RV measurement, however only 12 of these could be MG members if they had a different RV to that measured. (iii) If the non-MG sample were badly contaminated by unresolved binaries, their luminosities would be higher than expected for a single star, leading to a bias in the SED-fitting towards younger ages. But the SED-fitting ages of the non-MG sample are if anything, older than the MG sample. (iv) The 3D velocity dispersions of the MG and non-MG stars are still quite different (7.3 vs 18.0 km~s$^{-1}$ respectively) when objects with only a single RV measurement are removed. 

\subsection{BANYAN$\Sigma$}\label{sec:BANYAN}

Following the discussion in $\S$\ref{sec:MG_definition}, an alternative test for MG membership was carried out using the BANYAN$\Sigma$ MG membership probability tool \citep{2018a_Gagne} using the same set of MGs. If a star is assumed to be a MG member if it has $>80$ per cent probability of membership in any of the tested groups, then there are 86 MG members (38 per cent). This fraction rises to 42 per cent if instead, the highest membership probabilities of belonging to any of the MGs are summed for all 225 NLYS. The distribution of membership probabilities is very bimodal; most objects have either a very high probability of membership for one or more MGs or a very low membership probability in any MG. The BANYAN$\Sigma$ membership probabilities are listed in Table~\ref{table:input_data} if they are $>0$ per cent.

The fraction of NLYSs that belong to MGs ($>80$ per cent) from BANYAN$\Sigma$ is significantly lower than using just a kinematic criterion, because the BANYAN$\Sigma$ membership probability also uses spatial proximity to previously defined MG members (see $\S$\ref{sec:MG_definition}). There are 58 objects that pass the kinematic membership criterion (or at least are not rejected) but are $<80$ per cent likely to be members of any MG according to BANYAN$\Sigma$. In all these cases at least one of the spatial coordinates is $>3\sigma$ from the MG centroids. {\ The majority of these objects (37/58) are kinematically associated with MGs younger than 25 Myr; in particular 10 objects kinematically associated with the TW Hya MG and 9 with the Carina MG and 6 each with the 32 Orionis and Beta Pic MGs.}

There are also 8 objects with a BANYAN$\Sigma$ membership probability $>$ 80 per cent, but rejected as MG members by the kinematic criterion. These objects have kinematics that yield a chi-squared just above the 95 per cent rejection threshold (i.e. they have velocities that are somewhat beyond where we would accept them for MG membership), but they appear to be high probability MG members in the BANYAN$\Sigma$ framework, presumably because they are very close to the relevant MG $XYZ$ centroids. One of these objects is AB~Dor, the eponymous member of the AB Doradus MG, which is rejected by our kinematic criterion with 99.7 per cent confidence, because its space motion is not consistent with the MG velocity defined in BANYAN$\Sigma$.

If the kinematic criterion result and the BANYAN$\Sigma$ results are regarded as the upper and lower limits, then the genuine number of NLYS that belong to MGs is most likely between 40 and 60 per cent. Figures~\ref{figure:plot_BANYAN} and~\ref{figure:UVW_BANYAN} show the same plots as Figures~\ref{figure:plot} and~\ref{figure:UVW} from the BANYAN$\Sigma$ analysis. Whilst the trends amongst the NLYS from the BANYAN$\Sigma$ analysis are similar, one important difference between the BANYAN$\Sigma$ analysis results and those based solely on kinematic criteria is that there appears to be no age dependence in the fraction of stars assigned to MGs. One important difference between the BANYAN$\Sigma$ analysis results and those based solely on kinematic criteria is that there appears to be no age dependence in the fraction of stars assigned to MGs. The BANYAN$\Sigma$-assigned fraction of non-MG stars with $0.5<M/M_{\odot}<1.0$ and ages $<20$ Myr (36/65) is very similar to that for the older NLYS (67/107). Figure~\ref{figure:age_comparisons_BANYAN} shows that this is predominantly due to a lack of BANYAN$\Sigma$ matches to the youngest MGs.

\section{Discussion}\label{sec:discussion}

We have identified a sample of 225 low-mass, Li-rich NLYS from kinematically unbiased surveys within 100 pc, that should be younger than 125 Myr. About $50 \pm 10$ per cent of these NLYS can be identified with one or more known MGs in the solar neighbourhood, the uncertainty reflecting differing methodological approaches to assigning MG membership, rather than Poissonian uncertainties. Both the distribution of EW(Li) with colour and the ages inferred from SED-fitting suggest that the youngest stars ($<20$ Myr) in our sample are more likely to be part of MGs, but this conclusion rests on assigning MG membership using kinematics alone. The objects that cannot be assigned to MGs are kinematically ``hotter'' than the MG members, with higher overall velocity dispersions and some examples with velocities that are $20-40\,{\rm km\,s}^{-1}$ away from the bulk of the young objects.

That many young stars are not MG members has been noted before \citep[see][]{2012a_Shkolnik, 2015a_Binks, 2017a_Riedel}. Recently, \citet{2020a_Ujjwal} compiled Gaia DR2 astrometry for 890 objects identified in the `Catalog of Suspected Nearby Moving Group Stars' \citep{2017a_Riedel}, finding 279 (31 per cent) that were incompatible with membership of any of the 13 MGs analysed in the LACEwING membership probability code. However, this catalogue is kinematically biased, containing many stars that were found in searches for young stars based on parent samples that were kinematically consistent with MGs.  It is likely to be more incomplete for NLYS that are not MG members than is the kinematically unbiased sample considered here.

That about half of young, Li-rich, G- and K-type stars ($0.5 < M/M_{\odot}<1.0$) are found in MGs supports the idea that most, but perhaps not all, stars are born in groups that are kinematically coherent \citep{2003a_Lada} and that this kinematic coherence can survive up to $\sim 100$ Myr, \citep[for a similar recent observation see, e.g.][]{2019a_Kounkel}. Whether the fraction of NLYS in MGs is similar at lower masses awaits confirmation; \citet{2017b_Riedel} found around 70 per cent MG membership in a sample of young ($<50$ Myr) M-dwarfs, but this was from a kinematically biased parent sample. Searching for such objects based only on kinematic association with known MGs may only recover half the young population.

\subsection{A newly identified kinematic concentration amongst non-MG objects}\label{sec:new_groups}

One possibility to explain the high fraction of NLYS that cannot be assigned to a known MG, is that they belong to previously-unrecognised MGs. We find one object in the non-MG population, J03004686$-$3708018, that has been spectroscopically confirmed as a member of the Meingast~1 young stellar stream \citep{2020a_Arancibia-Silva}, with an RV and EW(Li) measurement entirely consistent with our compiled data. Figure~\ref{figure:UVW}~does suggest there may exist some small concentrations in velocity space that are not kinematically consistent with any of the known MGs within 100 pc. To identify over-densities in $UVW$ space the DBSCAN clustering algorithm \citep{1996a_Ester} was performed for the 89 NLYS that did not satisfy kinematic criteria of any of the MGs.  A Euclidean nearest-neighbour threshold of $1.7\,{\rm km\,s}^{-1}$ (typical of the 3D velocity dispersions of nearby MGs) and a minimum number of nearest neighbours of 4 was used. An advantage of DBSCAN is that kinematic substructures can be identified, regardless of their morphology.

DBSCAN identified one small kinematic grouping of 5 objects with a velocity centroid of $U = -8.2 \pm 0.2 \pm 0.4$, $V = -21.4 \pm 1.1 \pm 0.4$, $W = -5.3 \pm 0.5 \pm 0.1\,{\rm km\,s}^{-1}$ (the first uncertainty is the group dispersion and the second is the mean error in measurement), which is close to that of the young \citep[$15 \pm 3$ Myr,][]{2016a_Pecaut} association Lower Centaurus Crux (LCC) ($U=-7.8 \pm 2.7, V=-21.5 \pm 3.8, W=-6.2 \pm 1.8\,{\rm km\,s}^{-1}$), that is centred beyond 100 pc. These objects are identified in Figures~\ref{figure:plot}~and~\ref{figure:UVW}~as filled red squares. The Li content, position in the absolute colour-magnitude diagram and estimated ages from SED fitting are consistent with an age $\lesssim 20$ Myr and Figure~\ref{figure:plot}b shows four of these objects have sky positions similar to constituents of the LCC.

\cite{2018a_Goldman} identified a large MG within the LCC containing 1844 members (the LCC MG), 104 of which are closer than 100 pc. The distances to the 5 objects in our kinematic grouping might lie in the tails of the distance distribution of LCC MG members. The $XYZ$ components of the 5 co-moving objects are compatible with those amongst the LCC MG objects with $d<100$ pc, and may represent the near edge of this extended kinematic group.

\subsection{Young, Lithium-rich low mass stars}\label{sec:Li-rich}

In Figure~\ref{figure:plot}a there are 39 NLYS with $G-K_{\rm s}>2.5$ and EW(Li)$> 350$ m\AA. The presence of a strong Li signature in these cool K and M dwarfs suggests they are most likely younger than $\sim 25$ Myr. All 34 Li-rich NLYS that are matched to a MG by a kinematic criterion alone are matched to a MG with a probable age $<25$ Myr. That the youngest objects in our sample are matched to the youngest MGs provides some confidence in our analysis.

%In this section we focus on stars with  $G-K_{\rm s} > 2.5$ (spectral-types $>$K5) and EW(Li) $>350$ m\AA, and are likely younger than 25 Myr. There are 32 NLYS with a kinematic match to at least one MG in our chi-squared analysis. All but one have a kinematic match to a MG younger than $\sim 25$ Myr (J12264842$-$5215070 is exclusively matched to Carina ($\sim 40$ Myr)). This provides confidence in our analysis that the youngest objects in our sample are matched to the youngest MGs.

There are however 5 Li-rich NLYS that are not kinematically matched with any MG within 100 pc. We identify one star, J16435690$-$2508367, whose $U$ velocity is $> 20\,{\rm km\,s}^{-1}$ different from any known MG or nearby SFR. There are 2 independent, consistent RV measurements for this object, suggesting it is not a large amplitude RV variable. The remaining four have velocity components that all match within $2\sigma$ to either the $\rho$ Ophiuchus SFR or Sco-Cen association, that are centered beyond 100 pc. Only one of these, J13444279$-$6347495 is spatially close to either $\rho$ Ophiuchus or Sco-Cen (21 pc from the centre of LCC). The other 3 objects are more distant ($40 < d_{\rm min}/ {\rm pc} < 80$).

%however, they might still be members whose velocities lie in the tails of the MG velocity distributions.

\subsection{Origins of non-MG objects}\label{sec:origins_of_nonMG_stars}

It is unclear whether the significant fraction $(\simeq 50$ per cent) of NLYS that are not considered members of MGs share their origins with MG members or were subject to different formation mechanisms or dynamical histories. Their velocity dispersion is much higher than the MG stars and this is unlikely to be explained by binarity or measurement uncertainties. The $\sim 10$ per cent tail of the kinematically hottest NLYS in our sample are generally among the fastest stars in the Galactic disk population within the nearest few hundred pc \citep[see figures 22 and 24 in][]{2018a_Katz}. A number of possibilities can be considered:
\begin{enumerate}[label=(\roman*)]

\item Whilst we see no evidence for further concentrations in the kinematics of the non-MG NLYS, beyond those discussed in $\S$\ref{sec:new_groups}, it is possible that they belong to multiple as yet unrecognised MGs whose members are mostly outside our survey horizon \citep[e.g. part of the extensive young stellar streams that have been recently discovered in the local disc volume][]{2019a_Meingast, 2020a_Ratzenbock}. An argument against this possibility is that their kinematic dispersion is much higher than defined by the ensemble of known local MGs.

\item The objects could have formed in very small groups, and as a result have few comoving siblings \citep[e.g.][]{1996a_Feigelson}; but again some explanation of why they have hotter kinematics than other young stars in the solar vicinity would be required. It is also not clear why such stars may be systematically older than the MG members, although we note that this result depends on whether membership is defined solely using kinematic criteria.

\item The objects may have been ejected as a result of dynamical interactions in dense stellar aggregates or multiple systems \citep{1995a_Sterzik}. Given the large velocity differences, the birth environments could be at considerable distances from the Sun and nearby very young examples would be rare since there are no large, dense star forming regions within 100 pc. A variant of this is that rather than being dynamically ejected, these stars may have formed in the high velocity tails of molecular gas associated with larger star forming regions.
\end{enumerate}

\section{Summary}

We have assembled a kinematically unbiased sample of 225 young ($\lesssim 125$ Myr) low-mass stars within 100 pc of the Sun, using observations of lithium to confirm youth. As a result, these nearby, young, low-mass stars (NLYS) sample the full $0-125$ Myr age range for $0.5 < M/M_{\odot} < 1.0$, but are limited to just the youngest objects ($\lesssim 30$ Myr) at higher and lower masses.

We find that about $50 \pm 10$ per cent of the NLYS can be assigned to previously known MGs within 100 pc of the Sun. The uncertainty here represents whether assignment depends only on kinematics or is made more restrictive by demanding some spatial association too. The unassigned objects have significantly higher velocity dispersions than the MG members and there is marginal evidence that they are also systematically older. Five of the unassigned objects may belong to the Lower Centaurus Crux population that are found mostly just beyond 100 pc; there are also a further five objects with cool temperatures and very strong Li signatures, characteristic of very young PMS objects, that cannot readily be identified with any known MG or star forming region.

The kinematically ``hot'' NLYS that are not members of known MGs may have their origins in unrecognised MGs beyond the solar neighbourhood; star formation in low density environments; or they may have been ejected from star forming regions. Establishing which (if any) of these scenarios are in operation will require significantly more work to determine the ages of the kinematically dispersed NLYS population and to model their distributions in phase space from a variety of hypothesised birth sites and environments.

\section*{Acknowledgments}

ASB and RDJ acknowledge the financial support of the STFC. NJW acknowledges an STFC Ernest Rutherford Fellowship (grant number ST/M005569/1). An anonymous referee has made suggestions that significantly improved the content of this paper.

\appendix
\section{Results from the BANYAN$\Sigma$ analysis}\label{sec:appendix}

\begin{figure*}
\centering
\includegraphics[width=0.75\textwidth,angle=0]{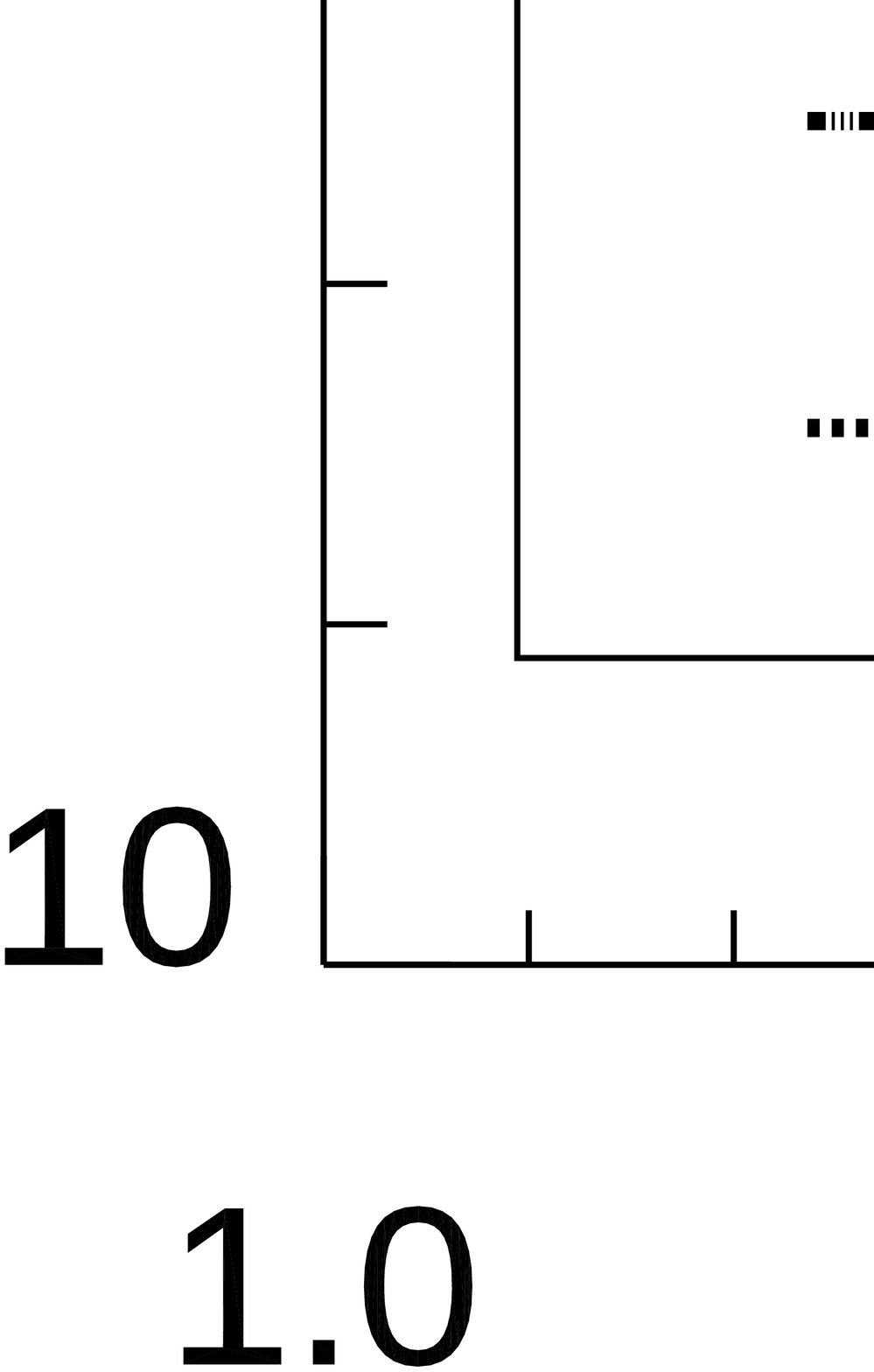}
\caption{The corresponding plot to Figure~\ref{figure:plot}, using the BANYAN$\Sigma$ results.}
\label{figure:plot_BANYAN}
\end{figure*}

\begin{figure*}
\centering
\includegraphics[width=0.75\textwidth,angle=0]{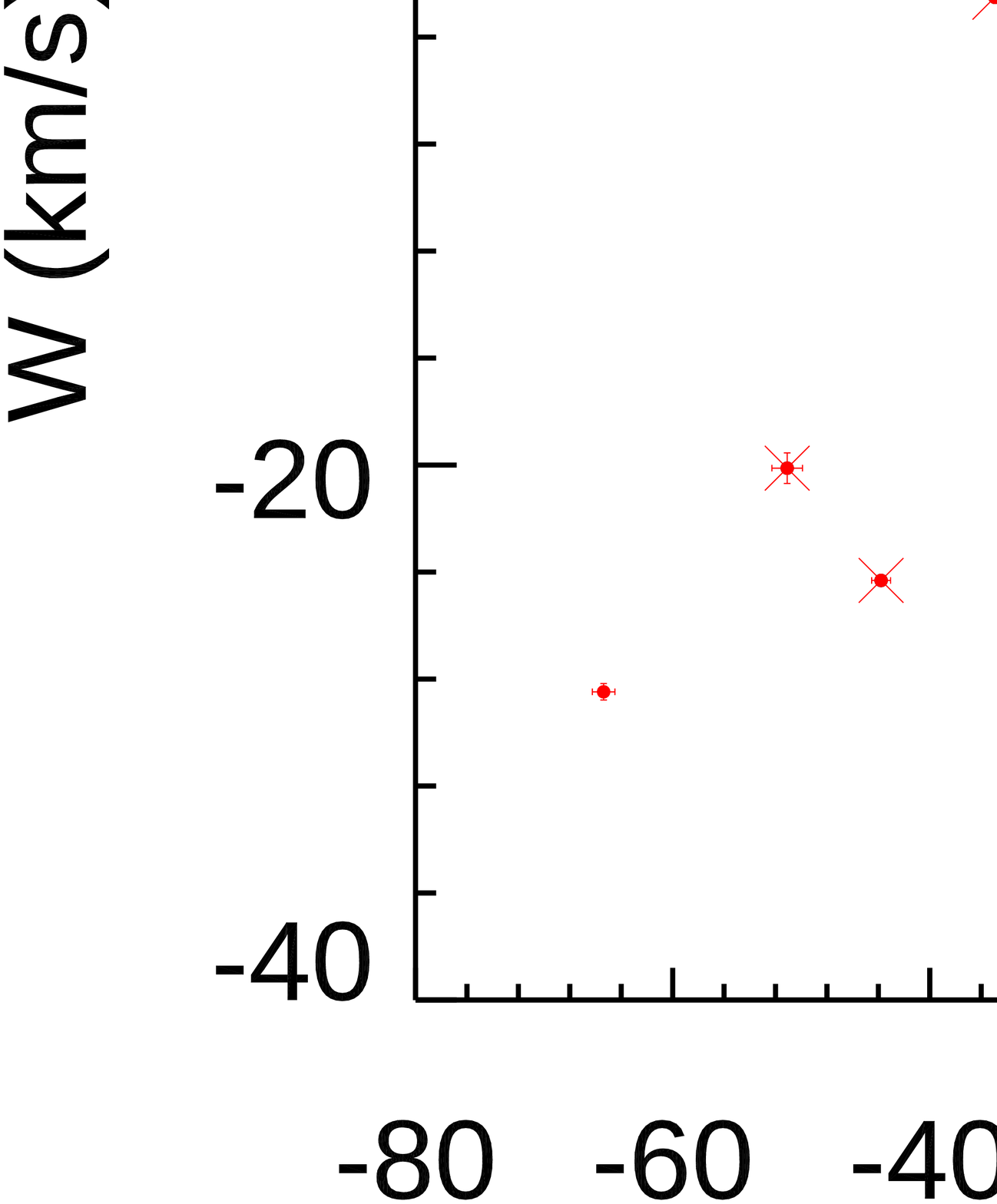}
\caption{The corresponding plot to Figure~\ref{figure:UVW}, using the BANYAN$\Sigma$ results.}
\label{figure:UVW_BANYAN}
\end{figure*}

\begin{figure}
\centering
\includegraphics[width=0.45\textwidth,angle=0]{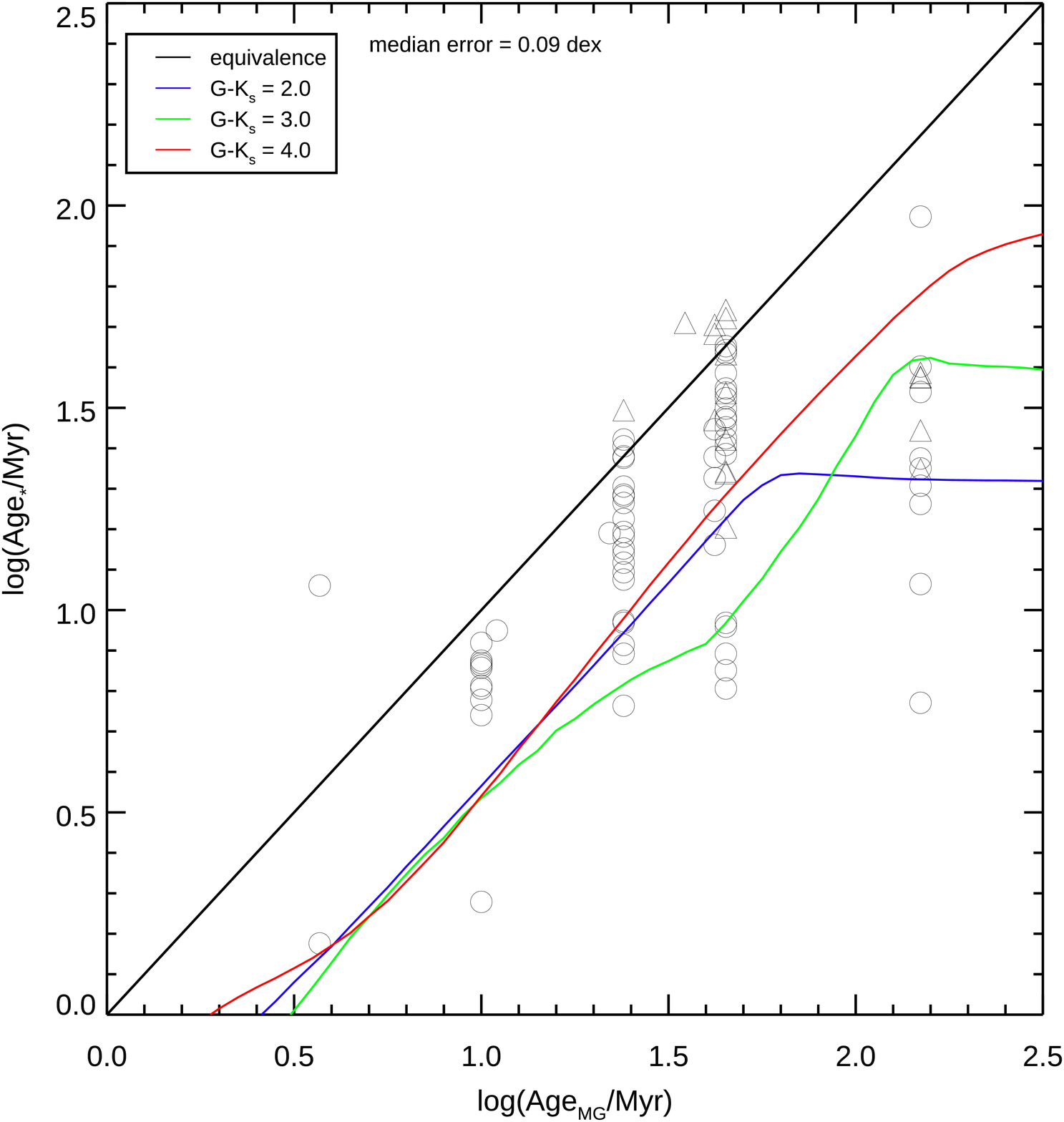}
\caption{The corresponding plot to Figure~\ref{figure:age_comparisons}, using the BANYAN$\Sigma$ results.}
\label{figure:age_comparisons_BANYAN}
\end{figure}

In $\S$\ref{sec:BANYAN} we describe our MG membership criteria using results from the BANYAN$\Sigma$ code. The plots presented here are the corresponding plots to Figures~\ref{figure:plot},~\ref{figure:UVW}~and~\ref{figure:age_comparisons}~using the BANYAN$\Sigma$ results. In Figures~\ref{figure:plot_BANYAN}~and~\ref{figure:UVW_BANYAN}~objects with a probability of MG membership $>80$ per cent are plotted as blue points. Figure~\ref{figure:age_comparisons_BANYAN}~shows the age comparison between SED fits and MGs for objects with $> 80$ per cent probability of MG membership in BANYAN$\Sigma$, where triangles denote age lower limits in the SED fits.

\bibliographystyle{mnras}
\bibliography{bibliography}
\label{lastpage}
\end{document}